\documentclass{templ}

\usepackage{cite}
\usepackage{booktabs}
\usepackage{threeparttable}

\hypersetup{pdftitle  = {A Harmonic-Regression ARMA model for rotating machinery dynamics identification and fault detection under varying conditions},
	pdfauthor = {D.M. Bourdalos, J.S. Sakellariou, S.D. Fassois}}

\title{A Harmonic-Regression ARMA model for rotating machinery dynamics identification and fault detection under varying conditions}

\author{D.M. Bourdalos}
\author{J.S. Sakellariou}
\author{S.D. Fassois}
  
\affil	{University of Patras, Department of Mechanical Engineering and Aeronautics \NewLineAffil 
Stochastic Mechanical Systems \& Automation (SMSA) Laboratory,\NewLineAffil
			26504 Patras, Greece \NewLineAffil
            e-mail: \textbf{sakj@upatras.gr}}

\date{}

\begin{document}



\abstract{
Fault detection in rotating machinery under varying operating conditions remains a challenging problem. Existing approaches are mainly based either on non-parametric vibration signal representations, parametric deep learning models, or on parametric ARMA-type models. While deep learning approaches often lack transparency and require large training datasets, ARMA-type models rely on the assumption of a purely rational spectrum, which can be restrictive for rotating machinery vibration signals exhibiting mixed spectral characteristics. To address these limitations, a novel Harmonic Regression ARMA (HR--ARMA) model is presented for rotating machinery dynamics identification, which combines an explicit harmonic regression part for representing the dominant deterministic cyclical components with an ARMA part for the remaining broadband stochastic dynamics. Based on this model and within a Multiple Model framework, fault detection is achieved under varying operating conditions. The HR--ARMA modelling performance is validated through simulation and experimental studies on a single-stage gearbox, demonstrating accurate representation of both cyclical and broadband dynamics, improved parsimony and significantly higher fault detection performance compared with conventional ARMA models, achieving at least 98\% True Positive Rate at 5\% False Positive Rate in the experimental study.}

\maketitle


\section{Introduction}
Rotating machinery, including gearboxes, bladed and reciprocating machines, are fundamental elements in numerous industrial, transportation, and energy applications. Faults in such components may lead to degraded performance, reduced availability, secondary damage, and, in severe cases, catastrophic failure \cite{bibliometric}. For this reason, condition-based maintenance has become increasingly important, as it enables maintenance actions to be scheduled according to the actual machine condition rather than fixed service intervals or failure occurrence. Among the available approaches, vibration-based condition monitoring has emerged as one of the most effective, owing to its high sensitivity, relatively low cost, simple instrumentation, and applicability without interrupting normal operation \cite{gearreview}.

Nevertheless, although numerous vibration-based fault detection methods have been developed, most studies assume constant operating conditions (OCs) \cite{bibliometric}. In practice, however, rotating machinery often operate under variable OCs, including different rotational speeds and loads, which may significantly affect the vibration response and obscure fault-related effects \cite{gearreview,helicoptercase}. Methods addressing this problem may be broadly classified into non-parametric and parametric, according to the representation adopted for the rotating machinery dynamics.

The largest part of the pertinent literature is based on non-parametric methods, employing frequency and time--frequency analysis, demodulation, cyclostationary analysis, and signal decomposition techniques, with the resulting representations used either directly for fault detection or for extracting features, often combined with machine-learning algorithms \cite{gryllias2021,MAURICIO2021,RANDALL2022paper,machines_features_new,Qi2022BlindIndicators,Civera2022SpectralEntropy,WANG2020107657}. Despite their effectiveness, these methods are often constrained by the suitability and tuning of the selected non-parametric representations and, in many cases, by the need for extensive datasets covering all health states considered.

To reduce the dependence on manually selected non-parametric representations and handcrafted features, more recent research has turned to parametric data-driven approaches, in which vibration signals are represented through model structures defined by a specific set of parameters estimated from data. Within this family, deep-learning models have formed a dominant line of work \cite{review_ai_mssp}, including convolutional and recurrent networks, autoencoder architectures, as well as transfer-learning, domain-adaptation, and domain-generalization schemes developed for fault detection under variable OCs \cite{Review_DeepLearning,SAE_4_A_novel_deep_autoencoder,AE4-SIGKRISI-speed-normalized,DTL1-adversarial,DTL3-domain-generalization-network}. In general, these methods seek to learn fault-discriminative representations that remain as invariant as possible across different OCs, thus reducing or eliminating the need for manual feature extraction. Despite their promising performance, however, they typically require large training datasets adequately covering the considered health states. Moreover, these models operate essentially as black-box schemes, without explicit correspondence to the underlying machinery dynamics, rendering their decisions difficult to interpret in physical terms and thus limiting their practical adoption. Although recent efforts toward explainable artificial intelligence are emerging, they remain at an early stage \cite{XAI_mssp}.

On the other hand, parametric statistical time-series (STS) methods provide an attractive alternative for rotating machinery fault detection under varying OCs. Rather than relying mainly on discriminative feature learning, these methods aim to represent the healthy machinery dynamics through structured stochastic models with a finite number of parameters, and subsequently assess the health state through statistically significant deviations from nominal model behavior. Their main advantages include compactness and a transparent relation between the model parameters and the underlying dynamics, thus supporting interpretable diagnostic decisions \cite{david2017,bourdalos_surv}. Typically, vibration signals acquired under healthy operation are used for the identification of multiple AR, vector AR and ARX models, usually one for each OC considered, while fault detection is based on the corresponding residuals via time-domain statistics or more specialized indices \cite{machinesEEDRIVEN,bourdalos2025iomac,bourdalosEWSHM,AR_different_speeds,AR_order_track_speed,ZHAN20071983,randal_ar_with_angular,YANG20105209,CHEN_LIN_VILIAM_MAKIS,var2021}. More recently Linear Parameter Varying (LPV)-type formulations, such as sparse LPV--AR, LPV--VAR, and LPV--ARMA models \cite{sparse2,sparse3,sparse4}, as well as Functionally Pooled formulations, with parameters expressed as explicit functions of the OCs such as FP--AR \cite{Bourdalos_mssp,bourdalos_surv} and vector FP--AR models \cite{bourdalos_machines_2026} have also been introduced. 

The widespread use of ARMA-type models in rotating machinery vibration analysis is not arbitrary, as AR models are well known to provide efficient representations of spectra with multiple peaks. This partly explains their broad adoption in the pertinent literature. At the same time, however, such models imply a purely continuous rational spectrum, which may be restrictive for rotating machinery vibration signals, since these typically contain deterministic cyclical components associated with shaft harmonics, gear meshing frequency harmonics, and modulation-induced sidebands, together with a broadband stochastic component related to the transmission path and structural dynamics. The resulting spectrum is therefore of mixed nature, with the cyclical spectral lines only indirectly approximated within a purely ARMA-type framework.

Motivated by the above mixed-spectrum nature of rotating machinery vibration signals, the present work introduces a novel Harmonic Regression ARMA (HR--ARMA) model for rotating machinery dynamics identification. The model is specifically designed to account for the mixed-spectrum nature of rotating machinery vibration signals by combining a Harmonic Regression part for the explicit representation of the dominant deterministic cyclical components with an ARMA part for the representation of the remaining broadband stochastic dynamics. Subsequently, a fault detection method is introduced by embedding the HR--ARMA model into the Multiple Model (MM) framework for properly accounting for varying operating conditions \cite{VAMVOUDAKIS2018}. 

The introduced model and the corresponding fault detection method are validated through both simulation and experimental studies. The simulation study offers a fully interpretable setting for systematically assessing the HR--ARMA model, including modelling accuracy, model structure selection, and its ability to capture both cyclical and broadband dynamics. The experimental study demonstrates the practical applicability and effectiveness of the introduced model and the fault detection methodology using hundreds of experiments on a single-stage spur gearbox operating under four rotational speeds and five load levels, while considering three levels of single-tooth gear crack. Comparative results against the conventional ARMA model are presented through scatter plots of the fault detection metric and Receiver Operating Characteristic (ROC) curves \cite{roc}.

\section{HR--ARMA model}
\subsection{HR--ARMA model structure}\label{subsec:HR_ARMA_model_structure}

Vibration signals from rotating machinery contain deterministic cyclical components mainly associated with shaft harmonics and, in the case of gearboxes, with the gear meshing frequency (GMF), its harmonics, and the corresponding sidebands. At the same time, the remaining part of the signal reflects broadband stochastic effects induced by the transmission path and the structural dynamics. As a result, such vibration signals may be viewed as possessing a mixed-spectrum structure, consisting of a discrete line spectrum superimposed on a continuous rational spectrum.

Accordingly, rotating machinery vibration signals may be well approximated as the superposition of (i) a finite deterministic cyclical part and (ii) a broadband stochastic part. Based on this, the Harmonic Regression AutoRegressive Moving Average (HR--ARMA) model is introduced. The model represents the mixed-spectrum dynamics through a cyclical Harmonic Regression part, which explicitly represents the dominant deterministic cyclical components, and a rational ARMA part, which represents the remaining broadband stochastic dynamics. The HR--ARMA model is defined  as follows:
\vspace{-4pt}
\begin{subequations}
\label{eq:HR_ARMA_model}
\begin{align}
y[n] &= \sum_{i=1}^{d} \tilde{A}_i \sin\!\left(2\pi f_i n + \tilde{\phi}_i\right)
+ \frac{c(\mathcal{B})}{a(\mathcal{B})} e[n]
\label{eq:HR_ARMA_model_a} \\
a(\mathcal{B}) &= 1 - \sum_{j=1}^{n_a} a_j \mathcal{B}^j,
\qquad
c(\mathcal{B}) = 1 - \sum_{j=1}^{n_c} c_j \mathcal{B}^j
\label{eq:HR_ARMA_model_c} \\
e[n] &\sim i.i.d.\,\mathcal{N}(0,\sigma_e^2)
\label{eq:HR_ARMA_model_d}
\end{align}
\end{subequations}

where $n=1,2,\ldots,N$ denotes the discrete angular-domain sample index, $y[n]$ the angularly resampled vibration signal, and $e[n]$ the model residuals, which, as indicated, constitute a Gaussian i.i.d.\ (independent and identically distributed) zero-mean process with variance $\sigma_e^2$. Moreover, $\tilde{A}_i$ and $\tilde{\phi}_i$ designate the amplitude and phase, respectively, of the $i$-th cyclical component, while $a(\mathcal{B})$ and $c(\mathcal{B})$ designate the AR and MA polynomials, respectively, with $\mathcal{B}$ denoting the backshift operator such that $\mathcal{B}^j y[n] := y[n-j]$. Finally, $f_i$ denotes the $i$th cyclical frequency (hereinafter referred to as cyclical order) normalized by the angular resampling frequency $f_s^{\theta}$, $d$ the number of cyclical components, and $n_a$, $n_c$ the AR and MA orders, respectively. It is noted that the model is defined in the angular domain and, therefore, the vibration signals should be first angularly resampled based on standard procedures \cite[pp. 148--151]{randall_book_all}. This is done in order to compensate for minor speed fluctuations, which may otherwise lead to spectral smearing of the deterministic cyclical components, even under nominally constant rotational speed \cite[p. 179]{randall_book_all}.

With proper manipulation, Eq.~\eqref{eq:HR_ARMA_model} may be written equivalently in the following form:
\vspace{-2pt}
\begin{equation}
\label{eq:HR_ARMA_model_equiv}
y[n]
=
\sum_{i=1}^{d}
\left[
A_i \sin(2\pi f_i n) + B_i \cos(2\pi f_i n)
\right]
+
\sum_{j=1}^{n_a} a_j y[n-j]
+
e[n]
-
\sum_{j=1}^{n_c} c_j e[n-j]
\end{equation}

\noindent with $e[n]$ as before. This representation is designated as an HR--ARMA$(d,n_a,n_c)$ model, with $n_a$ designating the AR order, $n_c$ the MA order, and $d$ the number of cyclical components. The constants $A_i$, $B_i$, and $f_i$ designate the coefficients and cyclical order of the $i$th cyclical component, respectively, while $a_j$ and $c_j$ designate the $j$th AR and MA parameters, respectively. For given cyclical orders $f_i$ and model residuals $e[n-j]$, Eq.~\eqref{eq:HR_ARMA_model_equiv} may be written in linear regression form with respect to the coefficients $A_i$, $B_i$, $a_j$ and $c_j$. The above are aggregated, together with the cyclical orders, in a single parameter vector $\boldsymbol{\theta}$ as follows ($^T$ designating transposition):
\vspace{-6pt}
\begin{align}\label{eq:param_vector}
\boldsymbol{\theta}=\big[f_1\ \ldots \ f_d \ \vdots \ A_{1} \ \ldots \ A_{d} \ \vdots \ B_{1} \ \ldots \ B_{d} \ \vdots \ a_{1} \ \ldots \ a_{n_a} \ \vdots \ c_{1} \ \ldots \ c_{n_c}\big]^T_{\big[3d+n_a+n_c\big]\times 1}
\end{align}

\subsection{HR--ARMA model estimation}
\label{subsec:HR_ARMA_estimation}

Estimation of the HR--ARMA$(d,n_a,n_c)$ model of Eq.~\eqref{eq:HR_ARMA_model_equiv}, that is, determination of the parameter vector $\boldsymbol{\theta}$ (Eq.~\eqref{eq:param_vector}) from a measured angular vibration signal $y[n]$, may be based on the corresponding model residuals. Specifically, based on Eq.~\eqref{eq:HR_ARMA_model_equiv}, the residuals corresponding to a candidate parameter vector $\boldsymbol{\theta}$ may be obtained as follows:
\begin{equation}
\label{eq:innovation_recursion}
e[n,\boldsymbol{\theta}]
=
y[n]
-
\sum_{i=1}^{d}
\left[
A_i \sin(2\pi f_i n)+B_i \cos(2\pi f_i n)
\right]
-
\sum_{j=1}^{n_a} a_j y[n-j]
+
\sum_{j=1}^{n_c} c_j e[n-j,\boldsymbol{\theta}]
\end{equation}
Model residuals are a function of the unknown parameter vector $\boldsymbol{\theta}$, and thus, model estimation is transformed into a non-quadratic optimization problem, leading to a Nonlinear least-squares optimization problem:
\begin{equation}
\label{eq:HR_ARMA_estimator}
\hat{\boldsymbol{\theta}}
=
\arg \min_{\boldsymbol{\theta}} 
\sum_{n=1}^{N} e^2[n,\boldsymbol{\theta}] 
\end{equation}

implemented via the trust-region-reflective algorithm (MATLAB function \texttt{lsqnonlin.m}). Since this nonlinear least-squares optimization is performed iteratively starting from an initial parameter vector, proper initialization of $\boldsymbol{\theta}$ may be achieved through the following steps.

\textbf{Step 1: Initialization of the cyclical orders.}
The initialization of the cyclical orders is based on the Periodogram Ratio Detection (PRD) method \cite{BerntsenPRD2022}. Specifically, two differently smoothed periodograms of the angular vibration signal $y[n]$ are estimated using two rectangular moving-average windows of lengths $N_1$ and $N_2$ ($N_1 \ll N_2$), and their ratio is computed as:
\begin{equation}
\label{eq:prd_statistic}
\mathrm{PR}[m]=\frac{P^{(1)}[m]}{P^{(2)}[m]}
\end{equation}

\noindent where $P^{(1)}[m]$ and $P^{(2)}[m]$ designate the corresponding smoothed periodograms obtained using $N_1$ and $N_2$, respectively. The initial cyclical-order set
\(
\widehat{\mathcal{F}}^{(0)}
=
[\hat{f}_1^{(0)},\ldots,\hat{f}_d^{(0)}]
\)
is then obtained by thresholding $\mathrm{PR}[m]$:
\begin{equation}
\label{eq:prd_metric}
\widehat{\mathcal{F}}^{(0)}
=
\left\{
m\,\Delta f_{\theta} \;\middle|\; \mathrm{PR}[m] > Th
\right\},
\qquad
\Delta f_{\theta}=\frac{f_s^{\theta}}{N}
\end{equation}
where $Th$ is a user-defined threshold, typically in the range $[8,15]$.

\textbf{Step 2: Initialization of the cyclical component coefficients.}
Once the initial cyclical-order set
\(
\widehat{\mathcal{F}}^{(0)}
=
[\hat{f}_1^{(0)},\ldots,\hat{f}_d^{(0)}]
\)
has been estimated, the corresponding coefficients $A_i$ and $B_i$ of Eq.~\eqref{eq:HR_ARMA_model_equiv} are initialized through ordinary least-squares (OLS) estimation of a simple Harmonic Regression model \cite{harmonic_on_tsa}:
\begin{equation}
\label{eq:initial_HR_model}
y[n]
=
\sum_{i=1}^{d}
\left[
A_i \sin(2\pi \hat{f}_i^{(0)} n)+ B_i \cos(2\pi \hat{f}_i^{(0)} n)
\right]
+r[n]
\end{equation}
This step provides the initial estimates
\(
\hat{A}_1^{(0)},\ldots,\hat{A}_d^{(0)}
\)
and
\(
\hat{B}_1^{(0)},\ldots,\hat{B}_d^{(0)}.
\)

\textbf{Step 3: Initialization of the ARMA parameters.}
Using the initial estimate of the cyclical component, the corresponding residual signal is obtained based on Eq. \eqref{eq:initial_HR_model} as:
\begin{equation}
\label{eq:preliminary_residual}
\hat{r}^{(0)}[n]
=
y[n]
-
\sum_{i=1}^{d}
\left[
\hat{A}_i^{(0)} \sin(2\pi \hat{f}_i^{(0)} n)+\hat{B}_i^{(0)} \cos(2\pi \hat{f}_i^{(0)} n)
\right]
\end{equation}
Subsequently, an ARMA$(n_a,n_c)$ model is estimated based on $\hat{r}^{(0)}[n]$ using standard model identification procedures \cite[pp. 199--201]{Ljung1999}, thus yielding the initial estimates of the AR and MA parameters:
\(
\hat{a}_1^{(0)},\ldots,\hat{a}_{n_a}^{(0)}
\)
and
\(
\hat{c}_1^{(0)},\ldots,\hat{c}_{n_c}^{(0)}.
\) The full initial parameter vector is then formed as:
\begin{equation}
\label{eq:initial_param_vector}
\widehat{\boldsymbol{\theta}}^{(0)}=
\Big[
\hat{f}_1^{(0)} \ \ldots \ \hat{f}_d^{(0)} \ \vdots \
\hat{A}_1^{(0)} \ \ldots \ \hat{A}_d^{(0)} \ \vdots \
\hat{B}_1^{(0)} \ \ldots \ \hat{B}_d^{(0)} \ \vdots \
\hat{a}_1^{(0)} \ \ldots \ \hat{a}_{n_a}^{(0)} \ \vdots \
\hat{c}_1^{(0)} \ \ldots \ \hat{c}_{n_c}^{(0)}
\Big]^T
\end{equation}
Starting from $\boldsymbol{\theta}^{(0)}$, all model parameters are jointly estimated through minimization of the criterion in Eq.~\eqref{eq:HR_ARMA_estimator}, thus leading to the final estimate $\hat{\boldsymbol{\theta}}$.

\section{Multiple HR--ARMA based fault detection under varying conditions}

A Multiple HR--ARMA based fault detection method is introduced for fault detection under varying OCs, based on the framework of Multiple Models (MM) \cite{VAMVOUDAKIS2018}. The method operates in two phases, namely the baseline (training) phase and the inspection phase.

\noindent \emph{Baseline (training) phase}: The concept behind the method is the representation of the healthy machinery dynamics via multiple HR--ARMA models (not necessarily of the same order), estimated using a number of $l$ healthy vibration signals with the machinery operating under different OCs in a range of interest. Thus, a set $\mathbb{M}_o$, with `o' indicating the `healthy' state, of HR--ARMA models is obtained using the estimation procedure of section~\ref{subsec:HR_ARMA_estimation}, with each model denoted as $M_{o,i}$ ($i=1,\ldots,l$).

\noindent \emph{Inspection phase}: Once an angular vibration signal $y_u[n]$ from the machinery under unknown health state and OC is available, it is driven through all models in the set $\mathbb{M}_o$. For each model $M_{o,i}$, the corresponding residual sequence is estimated based on Eq.~\eqref{eq:innovation_recursion} using the estimated parameter vector $\theta_{o,i}$ of the model $M_{o,i}$ as follows:
\begin{equation}
\hat{e}_{u_i}[n] = y_u[n]
-\sum_{r=1}^{d_i}\left[\hat{A}_{o,i,r}\sin(2\pi \hat{f}_{o,i,r} n)+\hat{B}_{o,i,r}\cos(2\pi \hat{f}_{o,i,r} n)\right]
-\sum_{j=1}^{n_{a,i}}\hat{a}_{o,i,j}y_u[n-j]
+\sum_{j=1}^{n_{c,i}}\hat{c}_{o,i,j}e_{u_i}[n-j]
\label{eq:MM_residual}
\end{equation}
\noindent where $d_i$, $n_{a,i}$, and $n_{c,i}$ denote the number of cyclical components, the AR order, and the MA order of model $M_{o,i}$, respectively, while $\hat{A}_{o,i,r}$, $\hat{B}_{o,i,r}$, $\hat{f}_{o,i,r}$, $\hat{a}_{o,i,j}$, and $\hat{c}_{o,i,j}$ designate the corresponding estimated parameters. The residual sequence $\hat{e}_{u_i}[n]$ is then used to assess whether the dynamics reflected in the tested signal $y_u[n]$ are consistent with those identified by the $i$-th model $M_{o,i}$ of the baseline set $\mathbb{M}_o$. In the present study, this consistency check is performed through a scalar metric selected as the $D$ statistic of the Pena--Rodriguez whiteness test \cite{PenaRodriguez2006}. Thus, letting $D_{u_i}$ denote the corresponding value obtained by passing the unknown signal through the baseline model $M_{o,i}$, the fault detection metric is defined as:
\begin{equation}
D=\min_{i=1,\ldots,l} D_{u_i}
\label{eq:MM_metric}
\end{equation}

Then, based on a user-defined threshold $L_{lim}$, fault detection is achieved as follows:
\begin{equation}
	\begin{gathered}
		D \leq L_{lim} \rightarrow \mbox{Healthy state}\\
		\mbox{Else} \hspace{15pt} \rightarrow \mbox{Faulty state}
	\end{gathered}
	\label{eq:MM}
\end{equation}


\section{Simulation Study}

The present simulation study offers a fully interpretable setting for systematically assessing the HR--ARMA model, including estimation accuracy, model structure selection, and its ability to capture both cyclical and broadband dynamics as well as to compare it with conventional ARMA modelling. To this end, the analytical signal simulation model presented in \cite{sparse4,sparse3} is employed to generate vibration signals from a single-stage gearbox. In the present study, however, this model is appropriately modified so that the transmission-path and structural vibrations are represented through a standard ARMA component, rather than through simple sinusoidal terms as in \cite{sparse4,sparse3}. This formulation is adopted because transmission-path and structural vibrations are more appropriately described as a broadband process with continuous spectral content, rather than deterministic
discrete-frequency component \cite[pp. 20--23]{randall_book_all}. The resulting simulation model is given by:
\vspace{-10pt}
\begin{equation}
\begin{aligned}
y[n] =\;& \underbrace{\sum_{k=1}^{K} A_k \sin\!\left(2\pi f_k n + \phi_k\right)}_{\text{shaft vibration}}
+\underbrace{\sum_{m=1}^{M} A_m[n]\sin\!\left(2\pi m f_g n + \phi_m\right)}_{\text{gear mesh vibration}} \\
&+\underbrace{\sum_{j=1}^{n_a} a_j\,y[n-j]-\sum_{j=1}^{n_c} c_j\,e[n-j]}_{\text{transmission path \& structural vibration}} + \underbrace{w[n]}_{\text{noise}}
\end{aligned}
\label{eq:simulation_model}
\end{equation}

\noindent where $y[n]$ denotes the simulated angular vibration signal. The first term represents shaft-related vibration components, with $K$ designating the number of retained shaft harmonics, $A_k$ and $\phi_k$ the amplitude and phase of the $k$th shaft harmonic, respectively, and $f_k$ its corresponding cyclical order, normalized by the angular sampling frequency $f_s^{\theta}$. The second term represents the gear mesh vibration, with $f_g$ designating the GMF order, $M$ the number of GMF harmonics and $\phi_m$ the corresponding constant phases. The amplitudes of the GMF harmonics are modulated via the function $A_m[n]$ which is expressed as: 

\begin{equation}
A_m[n] = A_{m_0} + A_{mod}\cos\!\left(2\pi f_{mod}n + \psi_{mod}\right)
\label{eq:Am_sim}
\end{equation}

\noindent where $A_{m_0}$ and $A_{mod}$ designate the mean and modulation amplitudes, respectively, $f_{mod}$ the modulation cyclical order, and $\psi_{mod}$ the corresponding phase. The third term represents the transmission-path and structural vibrations through an ARMA$(n_a,n_c)$ component, with $n_a$ and $n_c$ designating the AR and MA orders, respectively, $a_j$ and $c_j$ the corresponding AR and MA parameters, and $w[n]$ zero-mean Gaussian white-noise process with variance $\sigma_e^2$ representing the model's innovation sequence. Thus, based on the simulation model of Eq.~\eqref{eq:simulation_model}, an angular vibration signal is generated with angular sampling frequency $f_s^{\theta}=401$ samples/rev and length $N=32080$ samples corresponding to $n_{\mathrm{rev}}=80$ input shaft revolutions. The signal contains $d=14$ deterministic cyclical components, including shaft and GMF harmonics with the corresponding sidebands, while the stochastic component is represented through an ARMA$(4,4)$ model. All simulation parameters are provided in Table~\ref{tab:sim_params} of Appendix~\ref{app:sim_param}.

\begin{figure}[t]
\centering
\includegraphics[width=\linewidth]{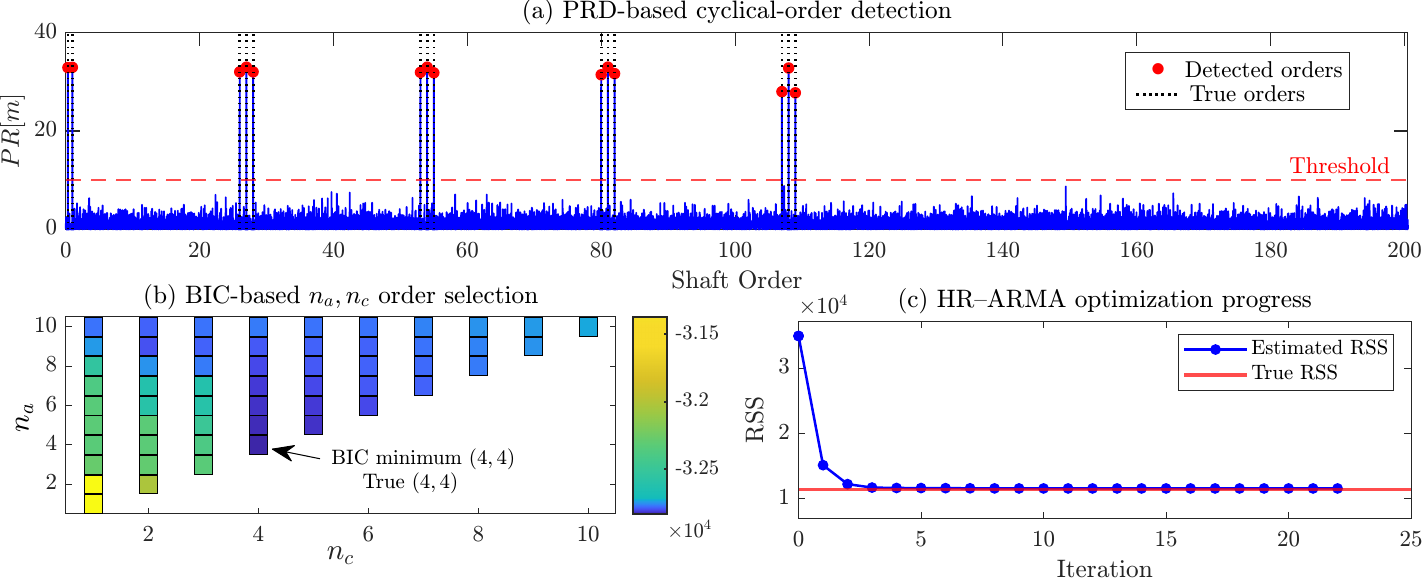}
\caption{HR--ARMA model estimation results in the simulation study. (a) PRD-based determination of the initial cyclical-order set $\widehat{\mathcal{F}}^{(0)}$, with the detected cyclical orders (red circles), the threshold (horizontal red dashed line), and the true cyclical orders (vertical black dashed lines); (b) BIC-based selection of the ARMA orders $(n_a,n_c)$ with the minimum correctly attained at the true values $(4,4)$; (c) progress of the nonlinear least-squares optimization for the HR--ARMA model estimation, shown in terms of the estimated cost function (RSS value) versus iteration number, together with the corresponding true RSS value.}
\label{fig:sim_fig1}
\end{figure}

Subsequently, an HR--ARMA model is estimated according to the procedure described in section~\ref{subsec:HR_ARMA_estimation}. More specifically, the initial cyclical-order set $\widehat{\mathcal{F}}^{(0)}$ is obtained through the PRD method based on Eq.~\eqref{eq:prd_metric}, as illustrated in Figure~\ref{fig:sim_fig1}(a). Then, based on this set, a harmonic-regression model of Eq.~\eqref{eq:initial_HR_model} is estimated via OLS, thus yielding the initial values $A_i^{(0)}$ and $B_i^{(0)}$ of the HR--ARMA cyclical-component coefficients $A_i$ and $B_i$. Subsequently, the ARMA orders $n_a$ and $n_c$ are determined based on the BIC criterion applied to the residuals of the harmonic regression model (Eq. \eqref{eq:preliminary_residual}), as shown in Figure~\ref{fig:sim_fig1}(b). As can be seen, the minimum BIC is achieved at the true values $n_a=n_c=4$. Then, a conventional ARMA$(4,4)$ model is estimated using the residual signal of Eq.~\eqref{eq:preliminary_residual}, and thus the AR and MA parameters of the HR--ARMA model are initialized. Finally, the complete HR--ARMA model is estimated through the nonlinear least-squares optimization of Eq.~\eqref{eq:HR_ARMA_estimator}, the progress of which is illustrated in Figure~\ref{fig:sim_fig1}(c) with respect to the iterations. As observed, the final estimated Residual Sum of Squares (RSS) converges correctly to the true value corresponding to the sum of squares of the simulation model innovation sequence.

\begin{figure}
\centering
\includegraphics[width=\linewidth]{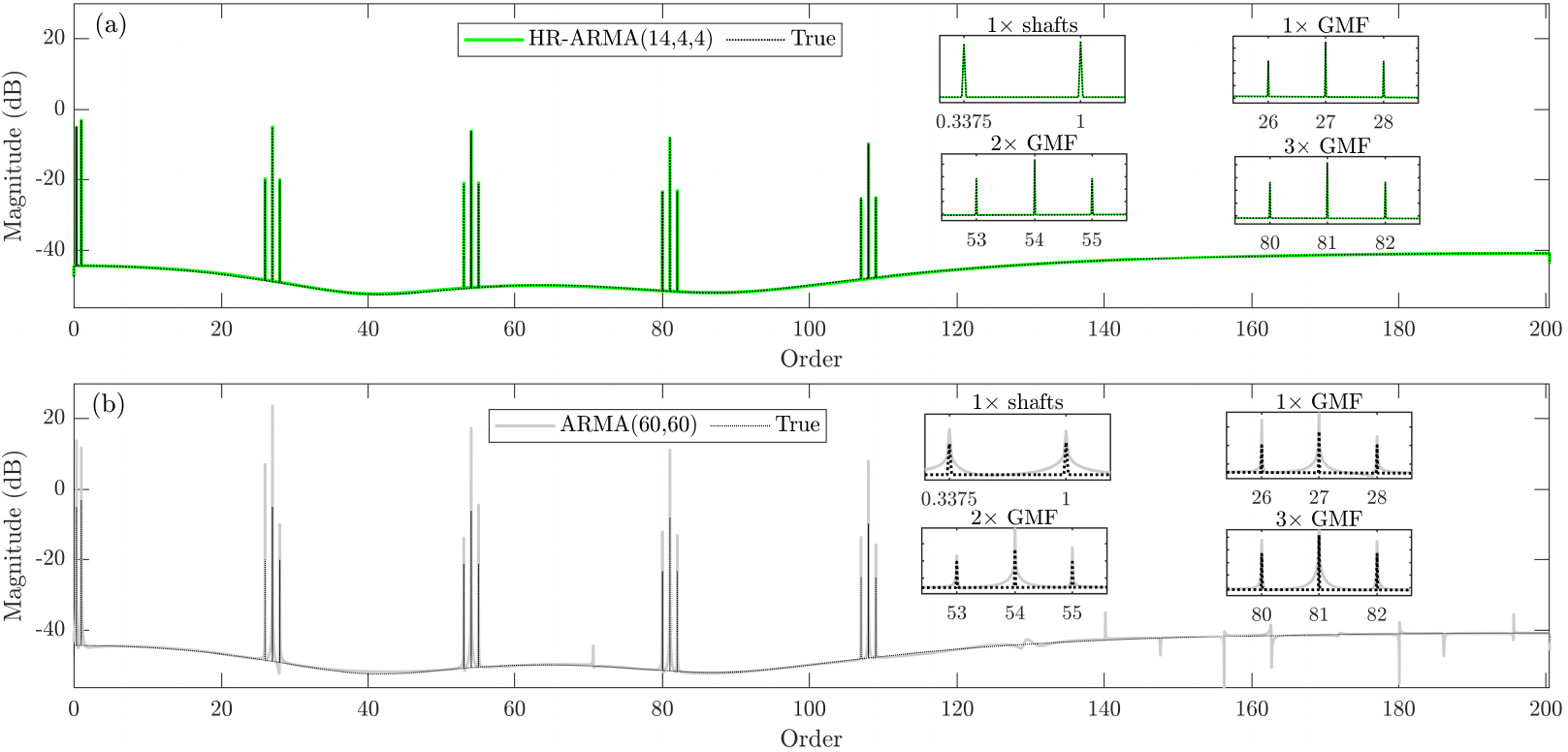}
\caption{Comparative modeling assessment via model-based mixed spectra in the simulation study. (a) true mixed spectrum (black dashed) versus the estimated HR--ARMA$(14,4,4)$--based mixed spectrum; (b) true mixed spectrum (black dashed) versus the ARMA$(60,60)$--based spectrum; the zoomed views focus on the shaft order and the first three GMF harmonics.}
\label{fig:sim_fig2}
\end{figure}

Thus, an HR--ARMA$(14,4,4)$ model is estimated for the simulated angular vibration signal (see all estimation details in Table \ref{tab:MODEL_DETAILS_simulation}). The corresponding HR--ARMA$(14,4,4)$--based mixed spectrum is presented in Figure~\ref{fig:sim_fig2}(a) with green color, where it is compared with the true theoretical mixed spectrum (black dashed) of the simulation model. As can be seen, a perfect agreement is observed over the entire order range, with the HR--ARMA model capturing accurately both the discrete line spectrum associated with the deterministic cyclical component and the continuous rational spectrum associated with the ARMA component. This agreement is further confirmed by the zoomed views around the shaft order and the first three GMF harmonics, where the true cyclical components and the corresponding sidebands are matched with excellent accuracy.

\begin{table}[b]
\centering
\caption{Simulation study - Details on the estimated models.}
\label{tab:MODEL_DETAILS_simulation}
\footnotesize
\renewcommand{\arraystretch}{1.18}
\setlength{\tabcolsep}{0pt}

\begin{tabular*}{\linewidth}{@{\extracolsep{\fill}}p{0.24\linewidth} p{0.16\linewidth} p{0.14\linewidth} p{0.16\linewidth} p{0.14\linewidth} p{0.12\linewidth}@{}}
\hline
\centering Estimated\\ model &
\centering Signal\\ length &
\centering No.\ of\\ parameters &
\centering Samples per\\ parameter &
\centering Final\\ RSS/SSS &
\centering No. of\\ iterations
\tabularnewline
\hline
\centering HR--ARMA$(14,4,4)$ &
\centering $N=32\,080$\\ samples &
\centering $50$ &
\centering $641.6$ &
\centering $15.43\%$ &
\centering $22$
\tabularnewline
\hline
\centering ARMA$(60,60)$ &
\centering $N=32\,080$\\ samples &
\centering $120$ &
\centering $267.3$ &
\centering $17.32\%$ &
\centering --
\tabularnewline
\hline
\end{tabular*}

\vspace{0.35em}

\begin{minipage}{\linewidth}
\scriptsize
\setlength{\parskip}{1.5pt}

\textit{Cyclical-order initialization.}
PRD method; threshold $Th=10$; moving-average window lengths $N_1=1$ and $N_2=33$.

\textit{Initialization of the ARMA parameters.}
Estimation using standard model identification procedures \cite[pp. 199--201]{Ljung1999}; Matlab function: \texttt{armax.m}.

\textit{Nonlinear least-squares optimization.} 
Trust-region-reflective algorithm; MATLAB function: \texttt{lsqnonlin.m}; maximum iterations $=15$; maximum function evaluations $=1.5\times10^{5}$; function tolerance $=10^{-10}$; step tolerance $=10^{-10}$.

\textit{Model validation.}
Verification of the model residual uncorrelatedness via the Auto Correlation Function (ACF) at risk level $\alpha=0.05$.
\end{minipage}
\end{table}

A comparison with the conventional ARMA model is also presented. In contrast to the ARMA part of the HR--ARMA model, which represents only the broadband stochastic component of the vibration signal, the conventional ARMA model is estimated directly from the angular vibration signal $y[n]$ and thus should represent both the deterministic cyclical components and the broadband stochastic component.
The comparison is conducted in terms of modeling accuracy with respect to the true mixed spectrum, with emphasis on the representation of shaft harmonics, GMF harmonics, and the corresponding sidebands, as well as on model parsimony. The ARMA model is estimated based on standard identification procedures \cite[pp. 199--201]{Ljung1999} using the same simulated signal. Full estimation details are provided in Table \ref{tab:MODEL_DETAILS_simulation}. The $n_a$ and $n_c$ orders are determined based on the BIC criterion as $n_a=n_c=60$. Figure~\ref{fig:sim_fig2}(b) shows the resulting ARMA$(60,60)$ model-based spectrum (grey color) with respect to the true mixed spectrum (black dashed). As shown this model captures reasonably well the broadband continuous part of the spectrum, but fails to reproduce accurately the amplitudes of the spectral lines associated with the deterministic cyclical component, as also indicated by the zoomed views around the shaft orders and the first three GMF harmonics. In addition, the ARMA spectrum exhibits artificial local peaks and valleys at higher orders, which do not correspond to true components of the simulated mixed spectrum, but rather constitute artifacts introduced by the high ARMA orders employed in an attempt to capture the deterministic line spectrum indirectly. Moreover, the model parsimony condition \cite[p. 492]{Ljung1999} is violated as a potentially unnecessarily large number of parameters is used to describe the system. In quantitative terms, the HR--ARMA$(14,4,4)$ model employs substantially fewer parameters, namely $50$ versus $120$ for the ARMA$(60,60)$ model, that is, 58\% fewer. It is also noted that the final RSS/SSS (RSS normalized by the series sum of squares) value of the conventional ARMA$(60,60)$ model is $17.32\%$, compared with $15.43\%$ of the HR--ARMA$(14,4,4)$ model.

\section{Experimental Study}

\subsection{The rotating machinery, the fault scenarios and the vibration signals}

\begin{figure}
\centering{\includegraphics[scale=0.63]{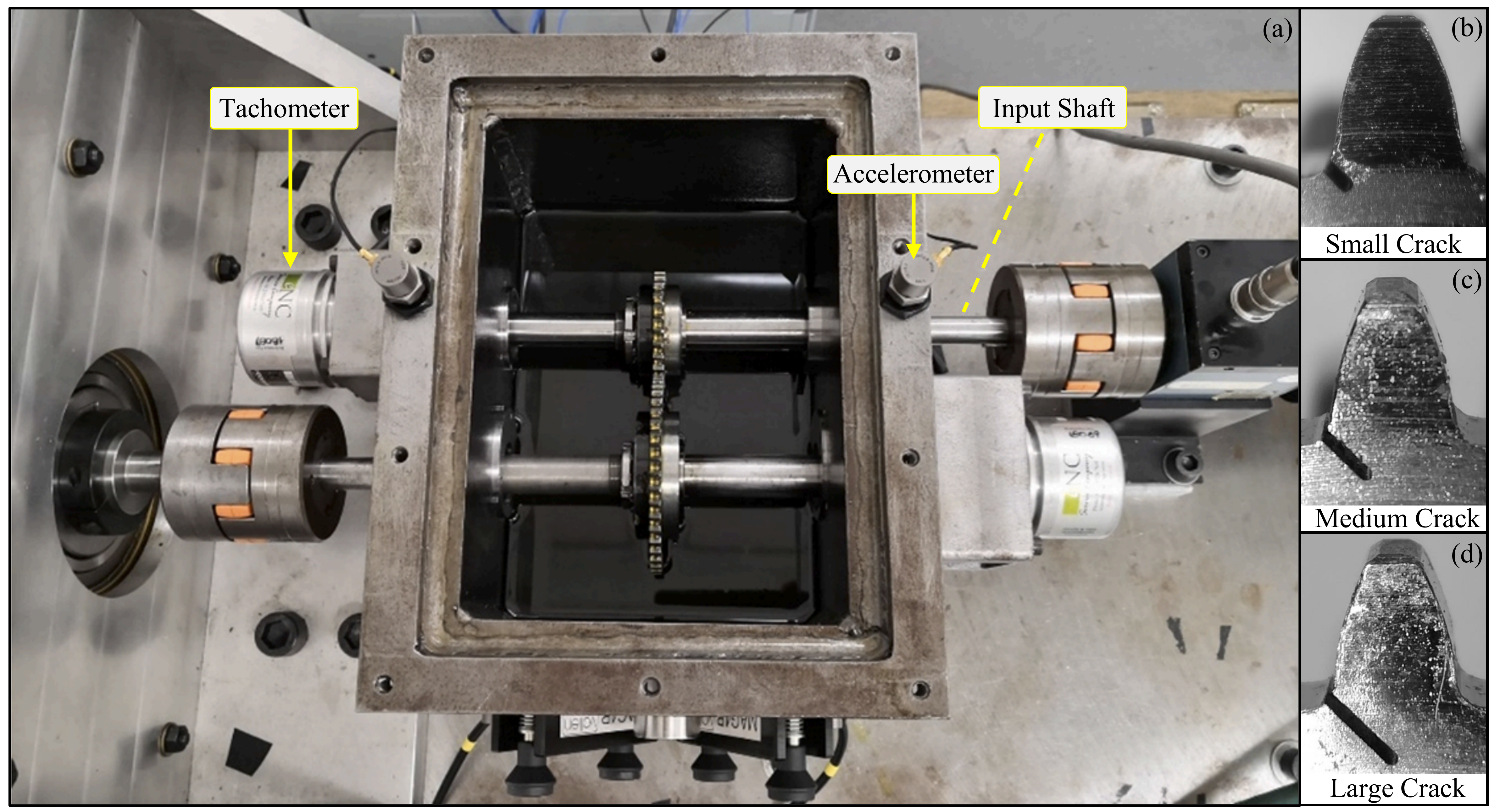}}
\caption{UNSW experimental set-up. (a) Photo of the gearbox including the sensors (accelerometer and tachometer) locations; the pinion single-tooth fault scenarios: (b) Small Crack, (c) Medium Crack and (d) Large Crack \cite{Randal_Data}.}
\label{fig:gearbox_UNSW}
\end{figure}

The experimental dataset employed in the present study has been collected by the Tribology and Machine Condition Monitoring Group of the University of New South Wales (UNSW), Sydney \cite{Randal_Data}, and is used to assess the introduced model and the corresponding fault detection method. The experimental test rig, shown in Figure~\ref{fig:gearbox_UNSW}(a), comprises a single-stage spur gearbox driven by an electric motor and loaded by means of an electromagnetic particle brake. The gearbox consists of a hunting-tooth steel gearset with a 27-tooth pinion and a 44-tooth driven gear. The drive motor operates at $4$ distinct rotational speeds within $\Omega \in [5,20]$~(rev/s), that is, $300$--$1200$~(rpm), with an increment of $5$~(rev/s) ($300$~rpm), regulated through a variable-frequency drive. In addition, five different load levels are considered, namely $0$, $5$, $10$, $15$, and $20$~(Nm). Three pinion cracks of different sizes, namely small (F1), medium (F2), and large (F3), are artificially introduced at $45^\circ$ from the fillet to the tooth centreline across the whole facewidth, as shown in Figure~\ref{fig:gearbox_UNSW}(b)--(d).

Vibration signals are acquired using a single uniaxial accelerometer mounted on the gearbox housing (see Figure~\ref{fig:gearbox_UNSW}(a)) at a sampling frequency of $f_s=200$~kHz. In addition, a tachometer is used to measure, synchronously, the rotational speed of the drive motor (see Figure~\ref{fig:gearbox_UNSW}(a)). Each original measurement available is segmented into 8 non-overlapping signals each of 2.5 seconds. Thus, a total of $640$ vibration signals are considered (see Table~\ref{tab:signals_details_UNSW}), spanning all health states and operating conditions.

\begin{table}[t]
\centering
\caption{Details on the vibration signals of the UNSW gearbox dataset.}
\label{tab:signals_details_UNSW}
\footnotesize
\setlength{\tabcolsep}{4pt}
\renewcommand{\arraystretch}{1.12}
\begin{threeparttable}
\begin{tabularx}{\textwidth}{
>{\raggedright\arraybackslash}X
*{4}{>{\centering\arraybackslash}X}
}
\toprule
Gearbox state &
\shortstack[c]{Rotational speed\\(rev/s)} &
\shortstack[c]{Load\\(Nm)} &
\shortstack[c]{No. of signals\\per speed \& load} &
\shortstack[c]{No. of signals\\per state} \\
\midrule

\multicolumn{5}{c}{\textbf{Training (learning) phase}} \\
\midrule
Healthy &
$\{5,10,15,20\}$ &
$\{0,5,10,15,20\}$ &
$1$ &
$20$ \\
\midrule

\multicolumn{5}{c}{\textbf{Inspection (testing) phase}} \\
\midrule
Healthy &
$\{5,10,15,20\}$ &
$\{0,5,10,15,20\}$ &
$7$ &
$140$ \\
Small crack &
-//- &
-//- &
$8$ &
$160$ \\
Medium crack &
-//- &
-//- &
$8$ &
$160$ \\
Large crack &
-//- &
-//- &
$8$ &
$160$ \\
\bottomrule
\end{tabularx}
\scriptsize

$^*$Sampling frequency: $f_s=200$ kHz; Time-domain signal length: $500\,000$ samples ($2.5$ s);
Angular resampling frequency: $f_s^{\theta}=990$ samples/rev; \\
Total No. of signals: $640$; training signals: $20$; inspection (test) signals: $620$.
\end{threeparttable}
\end{table}

\subsection{Experimental assessment of the HR-ARMA modeling}\label{subsection5.2}

\begin{figure}[t]
\includegraphics[width=\linewidth]{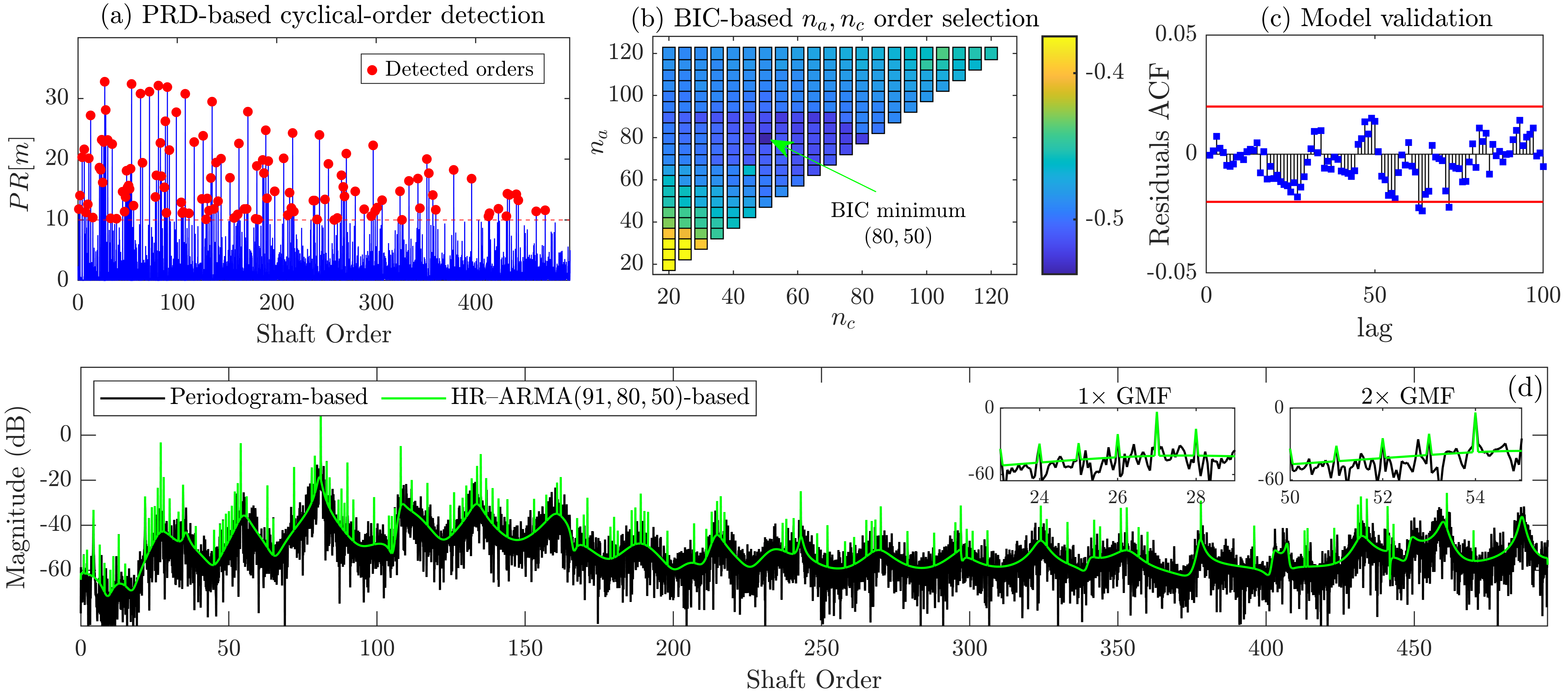}
\caption{HR--ARMA model estimation in the experimental study. (a) PRD-based determination of the initial cyclical-order set $\widehat{\mathcal{F}}^{(0)}$, with the detected cyclical orders (red circles) and the threshold (horizontal red dashed line); (b) BIC-based selection of the ARMA orders $(n_a,n_c)$ with the minimum shown; (c) HR--ARMA$(91,80,50)$ model validation via residual uncorrelatedness examination using the Auto Correlation Function (ACF) with the horizontal red lines denoting the statistical bounds at risk level of $\alpha=0.05$; (d) Comparison of the HR--ARMA($91,80,50$) model-based mixed spectrum (green) with the periodogram-based PSD estimate (black), including zoomed views around 1st and 2nd GMF.} 
\label{fig:EXP_Modeling}
\end{figure}

An indicative vibration signal corresponding to healthy state and rotational speed of $15$ (rev/s) and load $20$ (Nm) is selected in order to demonstrate the HR--ARMA modeling procedure in the experimental case. The model is estimated according to section~\ref{subsec:HR_ARMA_estimation}, including PRD-based initialization of the cyclical-order set (shown in Figure~\ref{fig:EXP_Modeling}(a)), OLS-based initialization of the cyclical component, BIC-based selection of the ARMA orders (Figure~\ref{fig:EXP_Modeling}(b)), initialization of the AR and MA parameters, and final joint estimation (see all estimation details in Table \ref{tab:MODEL_DETAILS_exp}). 

The above procedure leads to an HR--ARMA($91,80,50$) model, which is validated through the Auto Correlation Function (ACF) of the model residuals, as shown in Figure~\ref{fig:EXP_Modeling}(c), where the horizontal red lines denote the statistical confidence intervals at risk level $\alpha=0.05$. Moreover, Figure~\ref{fig:EXP_Modeling}(d) presents the comparison of the HR--ARMA($91,80,50$)-based mixed spectrum with the corresponding periodogram-based PSD estimate obtained directly from the angular vibration signal using a rectangular window. As observed, the model-based mixed spectrum is in good agreement with the periodogram-based PSD estimate, revealing that the model captures satisfactorily both the broadband rational PSD as well as the dominant spectral lines, including those around the first two GMF harmonics shown in the zoomed views.

\subsection{Experimental fault detection results}

\noindent \emph{Baseline (training) phase}: Based on $l=20$ vibration signals corresponding to the $20$ OCs considered (speeds and loads) as shown in Table~\ref{tab:signals_details_UNSW} and following the procedure described in section~\ref{subsec:HR_ARMA_estimation}, $20$ HR--ARMA models are estimated to form the $\mathbb{M}_o$ set of models, representing the healthy gearbox dynamics under varying OCs. 

\begin{figure}[t]
\centering{\includegraphics[scale=0.75]{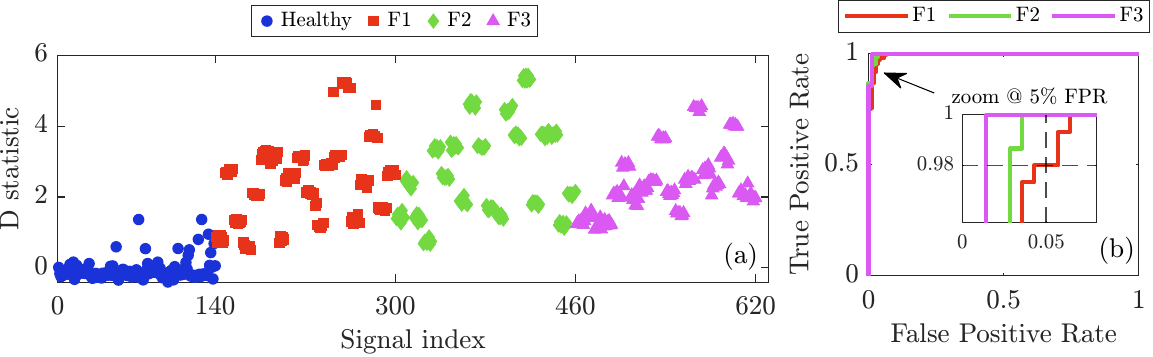}}
\caption{Experimental fault detection results. (a) Pena-Rodriguez $D$ statistic and, (b) corresponding ROC curves; 620 test signals in total (inspection phase) from the healthy and faulty gearbox under 4 different rotational speeds and 5 different loads.} 
\label{fig:FD_Results}
\end{figure}

\noindent \emph{Inspection phase}: 140 vibration signals from the healthy gearbox and 160 from each fault scenario are used in this phase for the method's fault detection performance assessment. The $D$ statistics of the multiple HR--ARMA based fault detection method for all considered inspection signals (620) are presented in Figure~\ref{fig:FD_Results} via scatter type plots and ROC curves. Based on the $D$-statistic values in Figure~\ref{fig:FD_Results}(a), the inspection signals corresponding to the healthy gearbox (blue) are observed to be almost fully separated from those corresponding to the considered fault scenarios, indicating overall very high detection performance. These conclusions are corroborated by the ROC curves in Figure~\ref{fig:FD_Results}(b). Specifically, the method attains $98\%$ TPR at $5\%$ FPR for the F1 case (small crack), while reaching $100\%$ TPR at the same FPR for the remaining fault scenarios (medium and large cracks).

\begin{table}[b]
\centering
\caption{Experimental study - Details on the estimated models.}
\label{tab:MODEL_DETAILS_exp}
\footnotesize
\renewcommand{\arraystretch}{1.18}
\setlength{\tabcolsep}{0pt}

\begin{tabular*}{\linewidth}{@{\extracolsep{\fill}}p{0.24\linewidth} p{0.16\linewidth} p{0.14\linewidth} p{0.16\linewidth} p{0.14\linewidth} p{0.12\linewidth}@{}}
\hline
\centering Estimated\\ model &
\centering Signal\\ length &
\centering No.\ of\\ parameters &
\centering Samples per\\ parameter &
\centering Final\\ RSS/SSS &
\centering No. of\\ iterations
\tabularnewline
\hline
\centering HR--ARMA$(91,80,50)$ &
\centering $N=29\,700$\\ samples &
\centering $403$ &
\centering $73.69$ &
\centering $8.33\%$ &
\centering $31$
\tabularnewline
\hline
\centering ARMA$(550,150)$ &
\centering $N=29\,700$\\ samples &
\centering $700$ &
\centering $42.43$ &
\centering $16.66\%$ &
\centering --
\tabularnewline
\hline
\end{tabular*}

\vspace{0.35em}

\begin{minipage}{\linewidth}
\scriptsize
\setlength{\parskip}{1.5pt}

\textit{Cyclical-order initialization.}
PRD method; threshold $Th=10$; moving-average window lengths $N_1=1$ and $N_2=33$.

\textit{Initialization of the ARMA parameters.}
Estimation using standard model identification procedures \cite[pp. 199--201]{Ljung1999}; Matlab function: \texttt{armax.m}.

\textit{Nonlinear least-squares optimization.} 
Trust-region-reflective algorithm; MATLAB function: \texttt{lsqnonlin.m}; maximum iterations $=100$; maximum function evaluations $=1.5\times10^{5}$; function tolerance $=10^{-6}$; step tolerance $=10^{-6}$.

\textit{Model validation.}
Verification of the model residual uncorrelatedness via the Auto Correlation Function (ACF) at risk level $\alpha=0.05$.
\end{minipage}
\end{table}

\subsection{Comparison with the conventional ARMA model}

\begin{figure}[h]
\includegraphics[width=\linewidth]{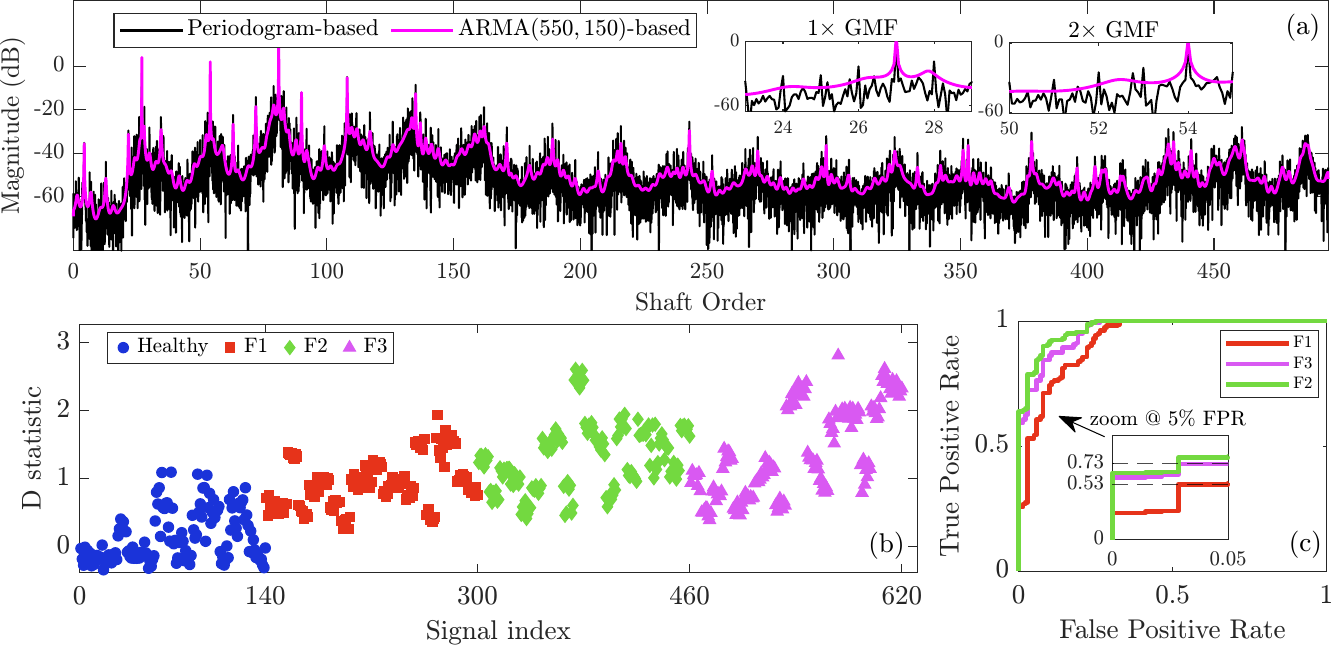}
\caption{(a) Comparison of the ARMA($550,150$) model--based spectrum (pink) with the periodogram-based PSD estimate (black), including zoomed views around 1st, 2nd and 3rd gear meshing frequencies (GMF); Experimental fault detection results using the conventional ARMA model: (b) Pena-Rodriguez $D$ statistic and, (c) corresponding ROC curves; 620 test signals in total (inspection phase) from the healthy and faulty gearbox under 4 different rotational speeds and 5 different loads.} 
\label{fig:ARMA_Results}
\end{figure}

A comparison with the conventional ARMA model is also performed in the experimental study, in order to further assess the advantages of the introduced HR--ARMA model. To this end, a standard ARMA model is estimated using the same healthy vibration signal considered in section~\ref{subsection5.2}, with the model orders selected based on the BIC criterion (see all estimation details in Table \ref{tab:MODEL_DETAILS_exp}). The resulting model is an ARMA$(550,150)$ model, validated through the ACF of model residuals, thus indicating that a high number of parameters is required to approximate the gearbox dynamics. In fact, the conventional HR--ARMA$(91,80,50)$ model employs approximately $42\%$ less parameters than the ARMA$(550,150)$ model. The corresponding model-based spectrum is presented in Figure~\ref{fig:ARMA_Results}(a), where it is compared against the periodogram-based PSD estimate obtained directly from the angular vibration signal. As observed, the conventional ARMA model captures reasonably well the broadband continuous part of the spectrum, but fails to reproduce with the same fidelity the dominant sharp spectral lines, including those around the first two GMF harmonics shown in the zoomed views. More specifically, only a limited number of dominant cyclical components, including certain GMF harmonics, are approximately captured, and even these exhibit noticeable amplitude mismatch, while several other components, including modulation sidebands, are not represented.

The effect of this reduced modeling accuracy is also reflected in the fault detection results. Figure~\ref{fig:ARMA_Results}(b) presents the corresponding $D$-statistic values, obtained by replacing the multiple HR--ARMA models with conventional ARMA models within the introduced fault detection method. As can be seen, a substantial overlap is now observed between the values corresponding to the healthy state (blue markers) and those corresponding to the faulty state for all considered fault scenarios. This reduced separability is further corroborated by the corresponding ROC curves in Figure~\ref{fig:ARMA_Results}(c) indicating poor fault detection performance with only $53\%$ TPR at $5\%$ FPR for the F1 case (small crack), and $73\%$ TPR at the same FPR for the remaining fault scenarios (medium and large cracks).

\section{Conclusions}
A novel HR--ARMA model has been introduced for rotating machinery dynamics identification. The introduced model accounts for the mixed-spectrum nature of rotating machinery vibration signals by combining a harmonic regression part, which explicitly represents the dominant deterministic cyclical components, including shaft harmonics, gear meshing frequencies, and the corresponding sidebands, with an ARMA part for the representation of the remaining broadband stochastic dynamics associated with the transmission path and the structural dynamics. Based on the introduced model and the concept of the Multiple Model framework, a fault detection method has also been developed for machinery operating under varying operating conditions. The performance of the introduced model has been validated, in terms of modelling accuracy and ability to capture both the cyclical and broadband dynamics, through both simulation and experimental studies on a single-stage gearbox. In both cases, the introduced model was shown to outperform the conventional ARMA alternative, while requiring a substantially smaller number of estimated parameters. The fault detection performance was validated through hundreds of experiments with the experimental gearbox operating under varying rotational speeds and loads and under three levels of single-tooth gear crack, achieving at least a $98\%$ True Positive Rate at a $5\%$ False Positive Rate, and clearly surpassing the conventional ARMA-based alternative, which exhibited inadequate performance.


\section*{Acknowledgments}

This research work was supported by the Hellenic Foundation for Research and Innovation (HFRI) under the 4th Call for HFRI PhD Fellowships (Fellowship Number 10820).



\bibliography{Refrences}

@Article{machinesEEDRIVEN,
AUTHOR = {Bourdalos, Dimitrios M. and Konstantinou, Xenofon D. and Koutsoupakis, Josef and Iliopoulos, Ilias A. and Kritikakos, Kyriakos and Karyofyllas, George and Spiliotopoulos, Panayotis E. and Saramantas, Ioannis E. and Sakellariou, John S. and Giagopoulos, Dimitrios and Fassois, Spilios D. and Seventekidis, Panagiotis and Natsiavas, Sotirios},
TITLE = {An AI Digital Platform for Fault Diagnosis and RUL Estimation in Drivetrain Systems Under Varying Operating Conditions},
JOURNAL = {Machines},
VOLUME = {14},
YEAR = {2026},
NUMBER = {1},
ARTICLE-NUMBER = {26}}

@article{bourdalosEWSHM,
author = {D. M. Bourdalos and I. A. Iliopoulos and J. S. Sakellariou},
journal = {Lecture Notes in Civil Engineering LNCE},
pages = {287-296},
title = {On the Detection of Incipient Faults in Rotating Machinery Under Different Operating Speeds Using Unsupervised Vibration-Based Statistical Time Series Methods},
volume = {254},
year = {2023}
}

@inproceedings{bourdalos2025iomac,
  author       = {Bourdalos, Dimitrios M. and Konstantinou, Xenofon D. and Sakellariou, John S. and Fassois, Spilios D.},
  title        = {Incipient gear fault detection under varying operating speed and load via {Multiple Model} vibration time series methods},
  booktitle    = {Proceedings of the 11th International Operational Modal Analysis Conference},
  year         = {2025},
  ADDRESS     = {Rennes, France},
  organization = {Institut National de Recherche en Informatique et en Automatique ({Inria})}}

@article{AR_order_track_speed,
year = {2019},
volume = {30},
number = {12},
pages = {125005},
author = {Wang, Lu and Xiang, Jiawei and Liu, Yi},
title = {A time–frequency-based maximum correlated kurtosis deconvolution approach for detecting bearing faults under variable speed conditions},
journal = {Measurement Science and Technology}}

@article{AR_different_speeds,
author = {McBain, Jordan and Timusk, Markus},
title = {Cross Correlation for Condition Monitoring of Variable Load and Speed Gearboxes},
journal = {Journal of Industrial Mathematics},
volume = {2014},
number = {1},
pages = {543056},
year = {2014}}

@article{RANDALL2022paper,
title = {A new angle-domain cepstral method for generalised gear diagnostics under constant and variable speed operation},
journal = {Mechanical Systems and Signal Processing},
volume = {178},
pages = {109313},
year = {2022},
issn = {0888-3270},
author = {R.B. Randall and W.A. Smith and P. Borghesani and Z. Peng}}

@article{WANG2020107657,
title = {Research on rolling bearing state health monitoring and life prediction based on PCA and Internet of things with multi-sensor},
journal = {Measurement},
volume = {157},
pages = {107657},
year = {2020},
author = {Heng Wang and Guangxian Ni and Jinhai Chen and Jiangming Qu}}

@article{Qi2022BlindIndicators,
  author       = {Qi, J. and Mauricio, A. and Gryllias, K.},
  title        = {Comparison of blind diagnostic indicators for condition monitoring of wind turbine gearbox bearings},
  journal = {Journal of Engineering for Gas Turbines and Power},
  year         = {2022},
  volume       = {144},
  number       = {4},
  pages        = {041019}}

@article{Civera2022SpectralEntropy,
  author       = {Civera, M. and Surace, C.},
  title        = {An application of instantaneous spectral entropy for the condition monitoring of wind turbines},
  journal = {Applied Sciences},
  year         = {2022},
  volume       = {12},
  number       = {3},
  pages        = {1059}}

@article{Randal_Data,
title = {Use of transmission error for a quantitative estimation of root-crack severity in gears},
journal = {Mechanical Systems and Signal Processing},
volume = {171},
pages = {108957},
year = {2022},
author = {Zhan Yie Chin and Pietro Borghesani and Yuanning Mao and Wade A. Smith and Robert B. Randall}}

@article{harmonic_on_tsa,
title = {Statistical modeling of gear vibration signals and its application to detecting and diagnosing gear faults},
journal = {Information Sciences},
volume = {259},
pages = {295-303},
year = {2014},
author = {Juliang Yin and Wenyi Wang and Zhihong Man and Suiyang Khoo}}

@article{bourdalos_machines_2026,
AUTHOR = {Bourdalos, Dimitrios M. and Sakellariou, John S.},
TITLE = {A Machine Learning Vibration-Based Methodology for Robust Detection and Severity Characterization of Gear Incipient Faults Under Variable Working Speed and Load},
JOURNAL = {Machines},
VOLUME = {14},
YEAR = {2026},
NUMBER = {1},
ARTICLE-NUMBER = {9}}

@article{BerntsenPRD2022,
title = {Periodogram ratio based automatic detection and removal of harmonics in time or angle domain},
journal = {Mechanical Systems and Signal Processing},
volume = {165},
pages = {108310},
year = {2022},
issn = {0888-3270},
author = {Jesper Berntsen and Anders Brandt},
}

@article{review_ai_mssp,
title = {Artificial intelligence for fault diagnosis of rotating machinery: A review},
journal = {Mechanical Systems and Signal Processing},
volume = {108},
pages = {33-47},
year = {2018},
author = {Ruonan Liu and Boyuan Yang and Enrico Zio and Xuefeng Chen}}

@article{Review_DeepLearning,
title = {A systematic literature review of deep learning for vibration-based fault diagnosis of critical rotating machinery: Limitations and challenges},
journal = {Journal of Sound and Vibration},
volume = {590},
pages = {118562},
year = {2024},
author = {Omri Matania and Itai Dattner and Jacob Bortman and Ron S. Kenett and Yisrael Parmet}}

@article{XAI_mssp,
title = {An explainable artificial intelligence approach for unsupervised fault detection and diagnosis in rotating machinery},
journal = {Mechanical Systems and Signal Processing},
volume = {163},
pages = {108105},
year = {2022},
author = {Lucas C. Brito and Gian Antonio Susto and Jorge N. Brito and Marcus A.V. Duarte}}

@article{SAE_4_A_novel_deep_autoencoder,
  author       = {Haidong Shao and Hongkai Jiang and Huiwei Zhao and Fuan Wang},
  title        = {A novel deep autoencoder feature learning method for rotating machinery fault diagnosis},
  journal      = {Mechanical Systems and Signal Processing},
  year         = {2017},
  volume       = {95},
  pages        = {187--204}}

@Article{machines_features_new,
AUTHOR = {Tayyab, Syed Muhammad and Chatterton, Steven and Pennacchi, Paolo},
TITLE = {Fault Detection and Severity Level Identification of Spiral Bevel Gears under Different Operating Conditions Using Artificial Intelligence Techniques},
JOURNAL = {Machines},
VOLUME = {9},
YEAR = {2021},
NUMBER = {8},
ARTICLE-NUMBER = {173}}

@Article{helicoptercase,
AUTHOR = {Hünemohr, David and Litzba, Jörg and Rahimi, Farid},
TITLE = {Usage Monitoring of Helicopter Gearboxes with ADS-B Flight Data},
JOURNAL = {Aerospace},
VOLUME = {9},
YEAR = {2022},
NUMBER = {11}}

@article{randal_ar_with_angular,
title = {Enhancement of autoregressive model based gear tooth fault detection technique by the use of minimum entropy deconvolution filter},
journal = {Mechanical Systems and Signal Processing},
volume = {21},
number = {2},
pages = {906-919},
year = {2007},
author = {H. Endo and R.B. Randall}}

@book{randall_book_all,
  title = {Vibration-Based Condition Monitoring: Industrial, Automotive and Aerospace Applications},
  shorttitle = {Vibration-Based Condition Monitoring},
  author = {Randall, Robert Bond},
  year = {2021},
  edition = {Second},
  publisher = {John Wiley \& Sons, Ltd},
  location = {Hoboken, NJ}}

@Article{PenaRodriguez2006,
  author  = {Pe\~na, Daniel and Rodr\'{\i}guez, Julio},
  title   = {The log of the determinant of the autocorrelation matrix for testing goodness of fit in time series},
  journal = {Journal of Statistical Planning and Inference},
  year    = {2006},
  volume  = {136},
  number  = {8},
  pages   = {2706--2718}
}

@article{Bourdalos_mssp,
title = {A statistical time series model-based method for robust detection of incipient faults in rotating machinery under different operating conditions},
journal = {Mechanical Systems and Signal Processing},
volume = {238},
pages = {113204},
year = {2025},
author = {D.M. Bourdalos and J.S. Sakellariou}}

@article{sparse2,
  author    = {Yuejian Chen and Stephan Schmidt and P. Stephan Heyns and Ming J. Zuo},
  title     = {A time series model-based method for gear tooth crack detection and severity assessment under random speed variation},
  journal   = {Mechanical Systems and Signal Processing},
  volume    = {156},
  pages     = {107605},
  year      = {2021}
}

@article{sparse3,
  author    = {Yuejian Chen and Ming J. Zuo},
  title     = {A sparse multivariate time series model-based fault detection method for gearboxes under variable speed condition},
  journal   = {Mechanical Systems and Signal Processing},
  volume    = {167},
  pages     = {108539},
  year      = {2022}
}

@article{sparse4,
  author    = {Yuejian Chen and Zihan Li and Yuan Jiang and Dao Gong and Kai Zhou},
  title     = {Sparse LPV-ARMA model for non-stationary vibration representation and its application on gearbox tooth crack detection under variable speed conditions},
  journal   = {Mechanical Systems and Signal Processing},
  volume    = {224},
  pages     = {112161},
  year      = {2025}
}

@article{AE4-SIGKRISI-speed-normalized,
  title={A speed normalized autoencoder for rotating machinery fault detection under varying speed conditions},
  author={Rao, Meng and Zuo, Ming J. and Tian, Zhigang},
  journal={Mechanical Systems and Signal Processing},
  volume={189},
  pages={109109},
  year={2023},
  publisher={Elsevier}
}

@article{DTL1-adversarial,
  title={Unknown working condition fault diagnosis of rotate machine without training sample based on local fault semantic attribute},
  author={Liu, Xuejun and Sun, Wei and Li, Hongkun and Li, Qiang and Ma, Zhenhui and Yang, Chen},
  journal={Advanced Engineering Informatics},
  volume={61},
  pages={102515},
  year={2024},
  publisher={Elsevier}
}

@article{DTL3-domain-generalization-network,
  title={Domain augmentation generalization network for real-time fault diagnosis under unseen working conditions},
  author={Shi, Yaowei and Deng, Aidong and Deng, Minqiang and Xu, Meng and Liu, Yang and Ding, Xue and Bian, Wenbin},
  journal={Reliability Engineering \& System Safety},
  volume={235},
  pages={109188},
  year={2023},
  publisher={Elsevier}
}

@book{Ljung1999,
  title={System Identification: Theory for the User},
  edition = {2nd edition},
  author={Ljung, L.},
  isbn={9780136566953},
  lccn={98018554},
  series={Prentice Hall information and system sciences series},
  year={1999},
  publisher={Prentice Hall PTR}
}

@inbook{roc,
author = {Duda, Richard and Hart, Peter and G.Stork, David},
year = {2001},
pages = {34--35},
title = {Pattern Classification},
isbn = {0-471-05669-3},
publisher = {John Wiley and Sons}
}

@article{gryllias2021,
title = {Enhanced bearing fault diagnosis using integral envelope spectrum from spectral coherence normalized with feature energy},
journal = {Measurement},
volume = {189},
pages = {110448},
year = {2022},
issn = {0263-2241},
author = {Bingyan Chen and Yao Cheng and Weihua Zhang and Fengshou Gu}
}

@article{MAURICIO2021,
title = {Cyclostationary-based Multiband Envelope Spectra Extraction for bearing diagnostics: The Combined Improved Envelope Spectrum},
journal = {Mechanical Systems and Signal Processing},
volume = {149},
pages = {107150},
year = {2021},
issn = {0888-3270},
author = {Alexandre Mauricio and Konstantinos Gryllias}
}

@INPROCEEDINGS{var2021,
  author={Li, Xin and Zuo, Hongfu and Hao, Pengcheng and Su, Ya and Liu, Haoyue and Xue, Cheng},
  booktitle={2021 Global Reliability and Prognostics and Health Management (PHM-Nanjing)}, 
  title={Early Fault Detection of Gearbox Using TSA and VAR Model Considering Load Variation}, 
  year={2021},
  volume={},
  number={},
  pages={1--6}}

@article{bibliometric,
  author={Chen, Jiayu and Lin, Cuiying and Peng, Di and Ge, Hongjuan},
  journal={IEEE Access}, 
  title={Fault Diagnosis of Rotating Machinery: A Review and Bibliometric Analysis}, 
  year={2020},
  volume={8},
  number={},
  pages={224985-225003}}

@article{gearreview,
author = {K. Feng and J. C. Ji and Q. Ni and M. Beer},
journal = {Mechanical Systems and Signal Processing},
pages = {109605},
title = {A review of vibration-based gear wear monitoring and prediction techniques},
volume = {182},
year = {2023},
unidentified = {p.}
}

@article{ZHAN20071983,
author = {Y. Zhan and C. K. Mechefske},
journal = {Mechanical Systems and Signal Processing},
number = {5},
pages = {1983-2011},
title = {Robust detection of gearbox deterioration using compromised autoregressive modeling and Kolmogorov-Smirnov test statistic. Part II: Experiment and application},
volume = {21},
year = {2007}
}

@article{YANG20105209,
author = {M. Yang and V. Makis},
journal = {Journal of Sound and Vibration},
number = {24},
pages = {5209-5221},
title = {ARX model-based gearbox fault detection and localization under varying load conditions},
volume = {329},
year = {2010}
}

@article{CHEN_LIN_VILIAM_MAKIS,
author = {C. Lin and V. Makis},
journal = {International Journal of Performability Engineering},
pages = {105},
title = {Application of Vector Time Series Modeling and T-squared Control Chart to Detect Early Gearbox Deterioration},
volume = {10},
year = {2014}
}

@article{VAMVOUDAKIS2018,
author = {K.J. Vamvoudakis-Stefanou and J.S. Sakellariou and S.D. Fassois},
journal = {Mechanical Systems and Signal Processing},
pages = {149-171},
title = {Vibration-based damage detection for a population of nominally identical structures: Unsupervised Multiple Model (MM) statistical time series type methods},
volume = {111},
year = {2018}
}

@article{david2017,
title = {Damage/fault diagnosis in an operating wind turbine under uncertainty via a vibration response Gaussian mixture random coefficient model based framework},
journal = {Mechanical Systems and Signal Processing},
volume = {91},
pages = {326-353},
year = {2017},
issn = {0888-3270},
author = {Luis David Avendaño-Valencia and Spilios D. Fassois}
}

@inproceedings{bourdalos_surv,
  TITLE = {{Vibration-based unsupervised detection of common faults in rotating machinery under varying operating speeds}},
  AUTHOR = {Bourdalos, Dimitrios M. and Sakellariou, John S.},
  BOOKTITLE = {{Surveillance, Vibrations, Shock and Noise}},
  ADDRESS = {Toulouse, France},
  ORGANIZATION = {{Institut Sup{\'e}rieur de l'A{\'e}ronautique et de l'Espace [ISAE-SUPAERO]}},
  HAL_LOCAL_REFERENCE = {78234},
  YEAR = {2023},
  MONTH = Jul,
  HAL_ID = {hal-04165675},
  HAL_VERSION = {v1}
}


	%
	%
	%



\appendix
\appendixheader

\section{Simulation Parameters \label{app:sim_param}}

\setcounter{table}{0}
\renewcommand{\thetable}{A\arabic{table}}

\begin{table}[h!]
\centering
\caption{Configuration parameters of the simulation study.}
\label{tab:sim_params}
\scriptsize
\renewcommand{\arraystretch}{1.1}
\setlength{\tabcolsep}{4pt}
\begin{tabular}{lll}
\toprule
\textbf{Parameter(s)} & \textbf{Description} & \textbf{Value} \\
\midrule

$f_s^\theta$ & Angular sampling frequency & $401$ (samples/rev) \\
$n_{\mathrm{rev}}$ & Number of revolutions & $80$ (rev) \\
$N$ & Total number of samples & $32080$ \\

$K$ & Number of shaft-related components & $2$ \\
$\{f_k\}_{k=1}^{K}$ & Shaft-related cyclical orders & $\{1,\;0.3375\}$ \\
$\{A_k\}_{k=1}^{K}$ & Shaft-related amplitudes & $\{1.0,\;0.8\}$ \\
$\{\phi_k\}_{k=1}^{K}$ & Shaft-related phases & $\{\pi/2,2\pi/3\}$ \\

$f_g$ & GMF order & $27$ \\
$M$ & Number of GMF harmonics & $4$ \\
$\{m f_g\}_{m=1}^{M}$ & GMF harmonic orders & $\{27,\;54,\;81,\;108\}$ \\
$\{\phi_m\}_{m=1}^{M}$ & GMF harmonic phases & $\{\pi/2,2\pi/3,\pi,\pi/4\}$ \\
$A_{m_0}$ & Mean GMF harmonic amplitudes & $\{0.80,\;0.70,\;0.55,\;0.45\}$ \\
$A_{\mathrm{mod}}$ & Modulation amplitude & $0.35\,A_{m_0}$ \\
$f_{\mathrm{mod}}$ & Modulation cyclical order & $1$ \\
$\psi_{\mathrm{mod}}$ & Modulation phase & $0$ \\

$n_a$ & AR order  & $4$ \\
$n_c$ & MA order  & $4$ \\
$\{a_j\}_{j=1}^{n_a}$ & AR parameters  & $\{1.10945,\;-0.63889,\;0.18978,\;-0.02359\}$ \\
$\{c_j\}_{j=1}^{n_c}$ & MA parameters & $\{1.61910,\;-1.66864,\;1.00377,\;-0.40909\}$ \\
$e[n]$ & Innovation sequence & $\mathcal{N}(0,\sigma_e^2)$ \\
$\sigma_e^2$ & Innovation variance & $1$ \\
\bottomrule
\end{tabular}
\end{table}

\end{document}